%% file: main.tex
\documentclass[sigconf]{acmart}

\usepackage{graphicx}
\usepackage{subcaption}
\usepackage{caption}

\AtBeginDocument{%
  \providecommand\BibTeX{{%
    \normalfont B\kern-0.5em{\scshape i\kern-0.25em b}\kern-0.8em\TeX}}}

\copyrightyear{2021}
\acmYear{2021}
\setcopyright{acmcopyright}\acmConference[LAK21]{LAK21: 11th International Learning Analytics and Knowledge Conference}{April 12--16, 2021}{Irvine, CA, USA}
\acmBooktitle{LAK21: 11th International Learning Analytics and Knowledge Conference (LAK21), April 12--16, 2021, Irvine, CA, USA}
\acmPrice{15.00}
\acmDOI{10.1145/3448139.3448188}
\acmISBN{978-1-4503-8935-8/21/04}

\begin{document}

\title{SAINT+: Integrating Temporal Features for EdNet Correctness Prediction}

\author{Dongmin Shin}
\affiliation{%
  \institution{Riiid! AI Research}
  \city{Seoul}
  \country{Republic of Korea}}
\email{dm.shin@riiid.co}

\author{Yugeun Shim}
\affiliation{%
  \institution{Riiid! AI Research}
  \city{Seoul}
  \country{Republic of Korea}}
\email{yugeun.shim@riiid.co}

\author{Hangyeol Yu}
\affiliation{%
  \institution{Riiid! AI Research}
  \city{Seoul}
  \country{Republic of Korea}}
\email{hangyeol.yu@riiid.co}

\author{Seewoo Lee}
\affiliation{%
  \institution{Riiid! AI Research}
  \city{Seoul}
  \country{Republic of Korea}}
\email{seewoo.lee@riiid.co}

\author{Byungsoo Kim}
\affiliation{%
  \institution{Riiid! AI Research}
  \city{Seoul}
  \country{Republic of Korea}}
\email{byungsoo.kim@riiid.co}

\author{Youngduck Choi}
\affiliation{%
  \institution{Riiid! AI Research}
  \city{Seoul}
  \country{Republic of Korea}}
\email{youngduck.choi@riiid.co}

\renewcommand{\shortauthors}{Dongmin, et al.}


\input{0_abstract}



\maketitle

\input{1_introduction}
\input{2_related_works}
\input{3_knowledge_tracing}
\input{4_SAINT+}
\input{5_experiments}
\input{6_conclusion}

\bibliographystyle{ACM-Reference-Format}
\bibliography{references}

\end{document}

%% file: 0_abstract.tex
\begin{abstract}
We propose SAINT+, a successor of SAINT which is a Transformer based knowledge tracing model that separately processes exercise information and student response information.
Following the architecture of SAINT, SAINT+ has an encoder-decoder structure where the encoder applies self-attention layers to a stream of exercise embeddings, and the decoder alternately applies self-attention layers and encoder-decoder attention layers to streams of response embeddings and encoder output.
Moreover, SAINT+ incorporates two temporal feature embeddings into the response embeddings: elapsed time, the time taken for a student to answer, and lag time, the time interval between adjacent learning activities.
We empirically evaluate the effectiveness of SAINT+ on EdNet, the largest publicly available benchmark dataset in the education domain.
Experimental results show that SAINT+ achieves state-of-the-art performance in knowledge tracing with an improvement of 1.25\% in area under receiver operating characteristic curve compared to SAINT, the current state-of-the-art model in EdNet dataset.
\end{abstract}

%% file: 1_introduction.tex
\section{Introduction}
The recent COVID-19 pandemic has raised needs for social distancing, leading many organizations to develop virtual and remote services to prevent widespread infection of the disease.
Accordingly, educational systems have developed remote learning environments including Massive Open Online Courses and Intelligent Tutoring Systems (ITSs).
Knowledge tracing, modeling a student’s knowledge state based on the history of their learning activities records, is a fundamental task for creating ITS that aims to provide personalized learning experiences to each student.
Traditionally, Bayesian Knowledge Tracing \cite{corbertt_1994,d2008more,yudelson2013individualized,pardos2010modeling,pardos2010navigating,pardos2013adapting,van2013properties,qiu2011does,kasurinen2009estimating,sao2013incorporating} and Collaborative Filtering based models \cite{thai2010recommender,lee2016machine} were common approaches for knowledge tracing.
However, with the widespread adoption of Deep Learning (DL) in many machine learning problems, DL based models \cite{piech_2015,lee_2019,liu_2019,zhang_2017,shen2020convolutional,vaswani_2017,pandy_2019,choi2020towards,ghosh2020context,nakagawa2019graph,tong2020hgkt} have become the de facto standard for knowledge tracing by capturing complex relations among interactions in students’ learning activities records.
The strength of the DL based models has become amplified with the advent of large scale public benchmark dataset in Artificial Intelligence in Education.
EdNet \cite{choi2020ednet} is such a publicly available dataset which is the largest in scale with more than 131M learning activities records from around 780K students.

In this paper, we propose SAINT+, a successor of SAINT \cite{choi2020towards} which enhances knowledge tracing with temporal feature embeddings, and empirically verify the effectiveness of the model on EdNet dataset.
SAINT is a Transformer \cite{vaswani_2017} based knowledge tracing model that separately processes information of exercise and student response.
Specifically, a stream of exercises that a student consumes is fed to an encoder, and a decoder gets a corresponding response sequence and encoder output sequence, computing the final output sequence of the model.
SAINT+ augments SAINT by integrating two temporal feature embeddings: elapsed time, the time taken for a student to answer, and lag time, the time interval between adjacent learning activities.
Empirical evaluations conducted on EdNet dataset show that SAINT+ improves SAINT, the current state-of-the-art knowledge tracing model on EdNet dataset, by 1.25\% in area under receiver operating characteristic curve (AUC).
Also, the experimental results show that incorporating the temporal features into the decoder input achieves the best AUC compared to incorporating them into the encoder input, and both the encoder and decoder input, verifying the hypothesis that separately processing exercise information and student response information is appropriate for knowledge tracing.

%% file: 2_related_works.tex
\section{Related Works}
Knowledge tracing is a fundamental task for many computer-aided educational applications and has been studied extensively in the field of AIEd.
Traditional approaches addressed knowledge tracing based on Bayesian Knowledge Tracing (BKT) \cite{corbertt_1994,d2008more,yudelson2013individualized,pardos2010modeling,pardos2010navigating,pardos2013adapting,van2013properties,qiu2011does,kasurinen2009estimating,sao2013incorporating} and Collaborative Filtering (CF) \cite{thai2010recommender,lee2016machine}.
Basically, BKT is the Hidden Markov Model where a latent variable represents evolving student knowledge.
BKT assumes the latent student knowledge as a set of binary variables: either the student mastered the knowledge or not.
Each latent variable is updated based on observations of the student correctly applying the knowledge which are also binary: either the student correctly or incorrectly answered a given exercise.
On the other hand, CF based approaches model students and exercises as low-rank matrices.
Each vector in the student matrix and exercise matrix represents latent traits of each student and latent knowledge required for each exercise, respectively.
The probability of a student correctly answers to an exercise is calculated by applying the sigmoid function to the dot product between the corresponding student and exercise vectors.

The advances of Deep Learning (DL) have given rise to neural network based knowledge tracing models.
DKT \cite{piech_2015} is the first DL based knowledge tracing model.
DKT models students’ evolving knowledge state through Recurrent Neural Network (RNN) which compresses their past learning activities in a hidden layer.
Like many RNN based models that commonly leverage attention mechanism, NPA \cite{lee_2019} models student knowledge through Bidirectional Long-Short Term Memory (Bi-LSTM) network equipped with an attention layer that weighs more importance to relevant parts of their learning history for prediction.
EKT \cite{liu_2019} is also a Bi-LSTM knowledge tracing model with an attention layer.
However, EKT addresses the cold start problem in knowledge tracing by exploiting not only students’ learning activities records but also text descriptions of exercises.
Not all DL based knowledge tracing models are based on RNN architecture.
DKVMN \cite{zhang_2017} is a memory-augmented neural network knowledge tracing model where the key matrix stores knowledge concepts and the value matrix stores students’ mastery levels of corresponding concepts.
CKT \cite{shen2020convolutional} is a knowledge tracing model that applies hierarchical convolutional operations to extract learning rate features from student’s learning activities history.
Applying self-attention mechanism in Transformer \cite{vaswani_2017} architecture, which is de facto standard to many sequential prediction tasks, to knowledge tracing is also an actively studied area.
SAKT \cite{pandy_2019} is the first knowledge tracing model with self-attention layers.
In each self-attention layer of SAKT, each query is an exercise embedding vector, and key and value are interaction embedding vectors.
SAINT \cite{choi2020towards} is the first Transformer based knowledge tracing model which leverages encoder-decoder architecture composed of stacked self-attention layers.
Unlike SAKT, SAINT gets separated streams of exercises and responses as inputs where a sequence of exercises are fed to the encoder, and a sequence of encoder outputs and responses are fed to the decoder.
AKT \cite{ghosh2020context} also adopts self-attention layers for knowledge tracing.
The attention weights in AKT are decayed exponentially based on the context-aware relative distance measure.
Moreover, AKT uses the Rasch model based exercise and exercise-response embeddings to avoid overparameterization and overfitting. Recently, several works \cite{nakagawa2019graph,tong2020hgkt} attempt to incorporate graph structure to the knowledge tracing model.
\cite{nakagawa2019graph} formulate knowledge tracing as a time series node-level classification task in graph structure and proposes GKT which extracts representation of each knowledge concept by aggregating representations of neighboring concepts.
HGKT \citep{tong2020hgkt} applies graph neural network to get hierarchical exercise graph which better represent groups of similar exercises.

%% file: 3_knowledge_tracing.tex
\section{Knowledge Tracing}
We formulate knowledge tracing as a task of predicting the probability of a student's answer being correct to a particular exercise given their previous interaction histories.
Formally, the student's learning activity is recorded as an interaction sequence $I_1, \dots, I_T$.
Each interaction $I_t = (E_t, R_t)$ is a tuple of \emph{exercise information} $E_t$, the $t$-th exercise given to the student with related metadata, such as the type of the exercise, and \emph{response information} $R_t$, the student's response to the exercise $E_t$ along with related metadata including the correctness of the response, the duration of time the student took to respond and the time interval between the current and previous interactions.
The student's response correctness $c_t \in \{0, 1\}$ is equal to 1 if the student answered the $t$-th exercise correctly and 0 if not. 
Thus, knowledge tracing aims to estimate the probability,
$$
\mathbb{P}[c_{t} = 1 | I_1, I_2, \dots, I_{t-1}, E_t].
$$

%% file: 4_SAINT+.tex
\section{SAINT+}

\begin{figure*}
    \centering
    \includegraphics[width=\textwidth]{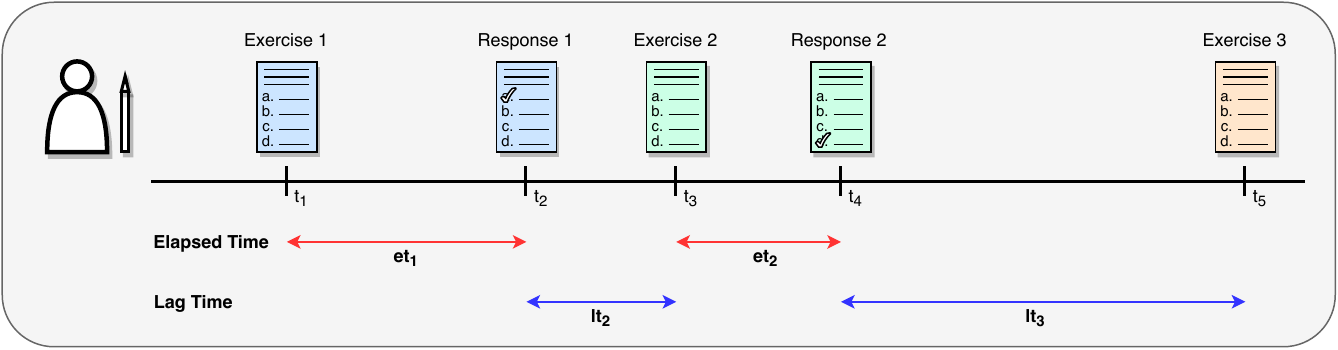}
    \caption{Descriptions of elapsed time and lag time.}
    \label{fig:time}
\end{figure*}

\subsection{SAINT: Separated Self-Attentive Neural Knowledge Tracing}
In this subsection, we give a brief review of SAINT, a \textbf{S}eparated Self-\textbf{A}ttent\textbf{I}ve \textbf{N}eural Knowledge \textbf{T}racing.
We refer the paper \cite{choi2020towards} for those who want to lean detailed aspects of SAINT.
SAINT is a knowledge tracing model based on Transformer \cite{vaswani_2017} architecture which consists of an encoder and a decoder.
It separates a stream of student interactions into two sequences: an exercise sequence and a response sequence.
Then, the encoder takes a sequence of exercise embeddings $E^e = [E_1^e, E_2^e, \ldots, E_T^e]$ as input and pass an output sequence $O^{enc}=[O^{enc}_1, O^{enc}_2, \ldots, O^{enc}_T]$ to the decoder.
The decoder additionally takes a shifted response embedding sequence $R^e = [S^e, R_1^e, R_2^e, \ldots, R_{T-1}^e]$ as input, whose first element is a start token embedding, to produce the final output sequence $\hat{c} = [\hat{c}_1, \hat{c}_2, \ldots, \hat{c}_{T}]$.
Each $\hat{c}_t$ is an estimated probability of the student's answer to the $t$-th exercise being correct given the current exercise information $E_t$ and the past interactions $I_1, I_2, \ldots, I_{t-1}$.

The most fundamental part of SAINT is a multi-head attention layer.
Let $h$ be the number of heads.
When input query matrix $Q$, key matrix $K$, and value matrix $V$ are given, we compute $Q_i = Q W_i^Q$, $K_i= K W_i^K$, and $V_i = V W^V_i$ for each $1 \leq i \leq h$, where $W_i^Q$, $W_i^K$, and $W_i^V$ are weight matrices of query, key and value, respectively.
Then, the scaled dot-product attention computes each head matrix, and the final output is a linear transformation of concatenated head matrices,
\begin{align*}
    \text{Multihead}(Q, K, V) 
    &= \text{Concat}(\text{head}_1, \text{head}_2, \ldots, \text{head}_h)W^O    \\
    \text{where} \; \text{head}_i 
    &= \text{Softmax}\Big( \text{Mask}\Big( \frac{Q_i K_i^T}{\sqrt{d}} \Big)\Big) V_i,
\end{align*}
where $d$ is a dimension of the query and key vectors, and $W^O$ is a weight matrix.
Note that the masking mechanism overwrites the region above the diagonal of the matrix $Q_iK_i^T$ with $-\infty$ so that
the corresponding region of the softmax output becomes zero.
This prevents the current position from attending to subsequent positions.
In other words, SAINT uses no future information from the sequence while training.

The encoder block consists of sequentially aligned $N$ copies of encoder layers.
A single encoder layer is a multi-headed self-attention layer with an upper triangular mask followed by a feed forward network (FFN) which is defined by 
\[
\text{FFN}(x) = \text{ReLU}(xW_1 + b_1)W_2 + b_2,
\] 
where $W_1, W_2$ and $b_1, b_2$ are weights and biases, respectively. 
Suppose $X$ is given as input to an encoder layer.
Then, the output sequence $O$ is computed as follows: 
\begin{align*}
    M &= X + \text{Multihead}(\text{LayerNorm}(X,X,X))\\  
    O &= M + \text{FFN}(\text{LayerNorm}(M)).
\end{align*}
Here, the input sequence $X$ for the foremost encoder layer is an exercise embedding sequence $E^e$ while each subsequent layer takes the feed forward output of the previous layer.
Note that we apply layer normalization \cite{ba2016layer} and skip connection \cite{he2016deep} to every sub-layer.

The decoder is a sequence of $N$ identical decoder layers, which consists of two multi-head attention layers with upper triangular masks followed by a feed forward network as well.
Suppose $X$ is an input sequence to a decoder layer.
If the layer is the foremost, $X$ is the response embedding sequence $R^e$.
Otherwise, it is the output sequence from the previous decoder layer.
The first layer is a multi-headed self-attention layer which only takes $X$.
Then, its output $M_1$ serves as queries for the second attention whose keys and values are the encoder output $O^{enc}$.
The computation can be summarized as follows:
\begin{align*}
    M_1 &= X + \text{Multihead}(\text{LayerNorm}(X,X,X))\\
    M_2 &= M_1 + \text{Multihead}(\text{LayerNorm}(M_1,O^{enc},O^{enc}))\\ 
    O &= M_2 + \text{FFN}(\text{LayerNorm}(M_2)).
\end{align*}
The output of the last decoder layer is passed to a fully connected layer to produce the final output of the model.

\subsection{Enhancing Knowledge Tracing with Temporal Feature Embeddings}

\begin{figure*}
    \centering
    \includegraphics[width=0.8\textwidth]{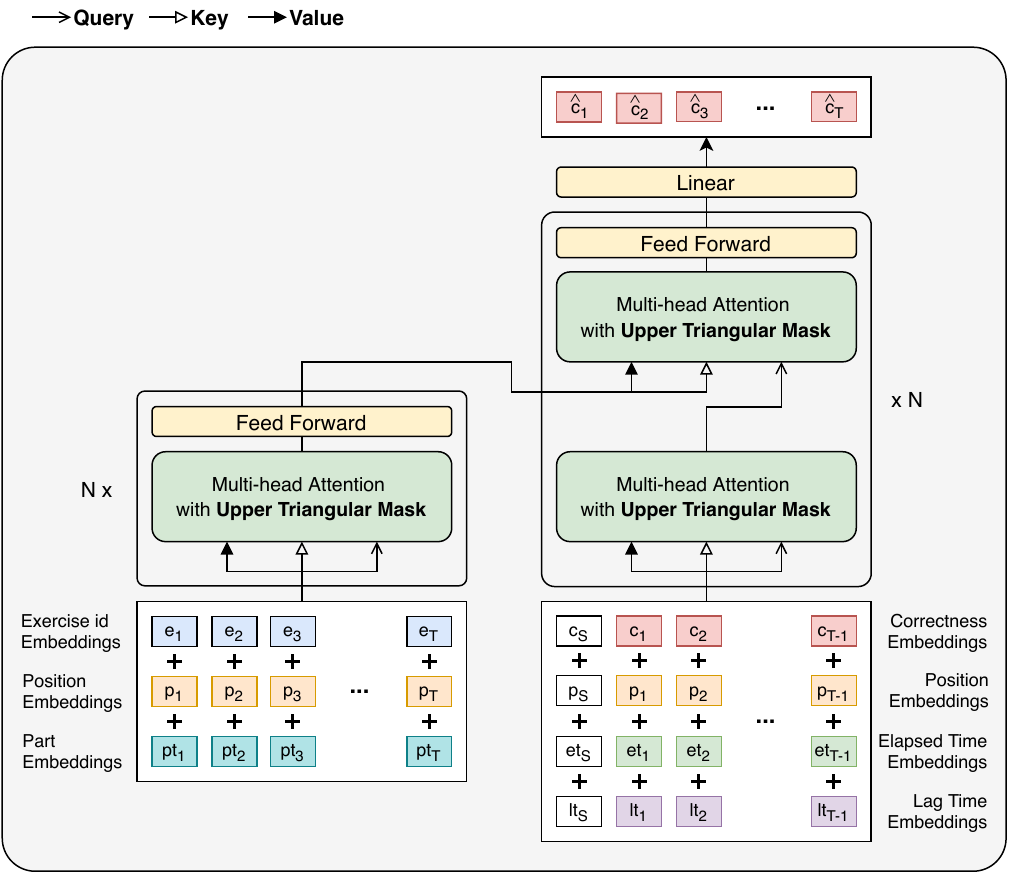}
    \caption{Model architecture of SAINT+.}
    \label{fig:model}
\end{figure*}

SAINT takes sequences of exercise embeddings and response embeddings, where each entry in the exercise embedding sequence is the sum of vectors of an exercise ID, an exercise category and the position, and each entry in the response embedding sequence is the sum of vectors of a response correctness and the position.
SAINT+ augments response embeddings with two temporal feature embeddings: elapsed time, the duration of time a student took to respond, and lag time, the time interval between the current and previous interactions (Figure \ref{fig:time}).
The embedding vectors for elapsed time and lag time are added to the response embeddings (Figure \ref{fig:model}).
In the following subsections, we provide detailed explanations of the two temporal feature embeddings.

\subsubsection{Elapsed Time}
Elapsed time is an amount of time that a student spent on solving a given exercise. 
For example, in Figure \ref{fig:time}, $\mathrm{et}_{1} = t_{2} - t_{1}$ (resp. $\mathrm{et}_{2} = t_{4} - t_{3}$) is the elapsed time for the exercise 1 (resp. exercise 2).
If the student does not have enough knowledge and skills for the exercise, it would be hard to respond correctly within the recommended time limit.
Hence, elapsed time provides strong evidence for a student's proficiency in knowledge and skills, and student's understanding of concepts associated with the exercise.

We propose two different approaches to embed elapsed times as latent vectors: continuous embedding and categorical embedding.
In continuous embedding, a latent embedding vector for an elapsed time $\mathrm{et}$ is computed as $\mathbf{v}_{\mathrm{et}} = \mathrm{et} \cdot \mathbf{w}_{\mathrm{elapsed\_time}}$, where $\mathbf{w}_{\mathrm{elapsed\_time}}$ is a single learnable vector.
For categorical embedding, unique latent vectors are assigned to each integer seconds.
We set the maximum elapsed time as 300 seconds and any time more than that is capped off to 300 seconds.

\subsubsection{Lag Time}
Lag time is the time gap between interactions, an important factor that affects complex phenomena occurring in students' learning process.
For instance, students tend to forget what they have learned as time passes.
If a lot of time passed after a student answers an exercise about certain concepts, it would be hard to respond to similar exercises correctly, even if they provided the correct answer to the exercise before.
On the other hand, students need time to refresh.
By taking a rest, their brains organize and arrange what they have learned, and prepare for the next learning session.
We define lag time as the time interval between the moment a student encountered the current exercise and the moment the student consumed the previous exercise.
For instance, in Figure \ref{fig:time}, lag time for the exercise 2 (resp. exercise 3) is $\mathrm{lt}_{2} = t_{3} - t_{2}$ (resp. $\mathrm{lt}_{3} = t_{5} - t_{4}$). 

Similar to the elapsed time embedding, we use continuous embedding and categorical embedding for lag time.
In continuous embedding, a latent embedding vector for a lag time $\mathrm{lt}$ is computed as $\mathbf{v}_{\mathrm{lt}} = \mathrm{lt} \cdot \mathbf{w}_{\mathrm{lag\_time}}$, where $\mathbf{w}_{\mathrm{lag\_time}}$ is a trainable vector.
For categorical embedding, lag times are discretized as integer minutes 0, 1, 2, 3, 4, 5, 10, 20, 30,  $\dots$, 1440. As a result, there are a total of 150 trainable unique latent vectors assigned to each integer minute.

%% file: 5_experiments.tex
\section{Experiments} \label{sec:exp}
\subsection{Dataset}
\begin{figure*}
    \centering
    \begin{subfigure}{0.4\textwidth}
        \includegraphics[width=\textwidth]{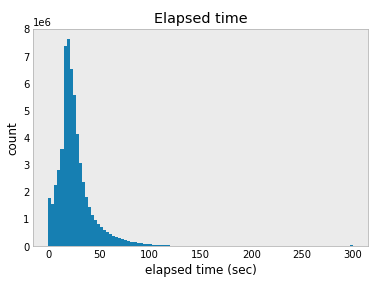}
    \end{subfigure}
    \begin{subfigure}{0.4\textwidth}
        \includegraphics[width=\textwidth]{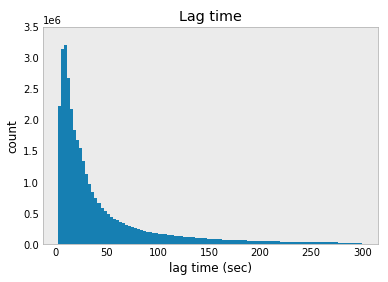}
    \end{subfigure}
    \caption{Distributions of elapsed time and lag time in an updated version of EdNet-KT1.}
    \label{fig:time_hist}
\end{figure*}

EdNet \cite{choi2020ednet} is the largest publicly available benchmark dataset in education domain consisting of interaction logs collected by Santa\footnote{\url{https://aitutorsanta.com}}.
We conduct experiments on an updated version of EdNet-KT1, which contains problem-solving logs from January 1st, 2019 to June 1st, 2020.
Details of the dataset statistics are provided in Table \ref{tab:ednet}.
Also, the distributions of elapsed time and lag time of the dataset are shown in Figure \ref{fig:time_hist}.
We use interaction logs of the most recent 100K students as test set and 80\% (resp. 20\%) of the remaining dataset are used as training (resp. validation) set. 

\begin{table}[ht]
\caption{Statistics of an updated version of EdNet-KT1 dataset.}
\centering
\begin{tabular}{l|c}\toprule
\# of interactions & 60294498 \\
\# of students & 678128 \\
\# of exercises & 14418 \\
\bottomrule
\end{tabular}
\label{tab:ednet}
\end{table}

\subsection{Baseline Models}
We compare SAINT+ against benchmark knowledge tracing models, DKT \cite{piech_2015}, DKVMN \cite{zhang_2017}, SAKT \cite{pandy_2019} and SAINT \cite{choi2020towards}:
\begin{itemize}
    \item DKT is a simple RNN-based model that uses concept and response correctness as input features, and models student's knowledge status as RNN's hidden state vectors.
    We choose an LSTM \cite{hochreiter1997long} architecture.
    Also, we consider each unique exercise id as a concept associated with the exercise.
    \item DKVMN is a memory augmented neural network based model where the key matrix stores knowledge concepts and the value matrix stores students’ mastery levels of the corresponding concepts.
    We use each exercise id as a concept of the corresponding exercise.
    \item SAKT is the first knowledge tracing model that utilizes Transformer's self-attention architecture. 
    It is a single encoder-based model where exercise ids are used as queries and interaction ids are used as keys and values. 
    \item SAINT is the first Transformer-based knowledge tracing model which leverages an encoder-decoder structure to process information of question and student response separately.
    The encoder takes a sequence of question embeddings and the decoder gets a sequence of encoder outputs and response embeddings to compute the final output.
\end{itemize}

\subsection{Model Training and Evaluation}
We use the accuracy (ACC) and the area under the receiver operating characteristic curve (AUC) as the performance metric. We pick the weight with the best validation AUC and evaluate them on the test set.
For SAINT+ and SAINT, we use Adam optimizer with $lr=0.001, \beta_1 = 0.9$, $\beta_2 = 0.999$ and $epsilon=1e-8$, and set warmup steps to 4000.
The window size, number of layers, dimension of the model, dropout rate and batch size is set to 100, 4, 512, 0.0 and 64, respectively.
For other benchmark models, both embedding and hidden dimensions of DKT, DKVMN and SAKT are searched over [50, 100, 150, 200, 256, 512], and the best results are reported on the test dataset.
Also, we set the number of latent concepts as 64 for DKVMN. 

\subsection{Main Results}

Table \ref{tab:main} shows the performance comparison of SAINT+ with the benchmark knowledge tracing models.
Compared to the benchmark models, SAINT+ achieves increases of ACC and AUC maximally up to 2.72\% and 3.61\%, respectively.
Also, SAINT+ improves SAINT with 1.03\% and 1.25\% gain in ACC and AUC, respectively, demonstrating the effectiveness of integrating the temporal features for knowledge tracing. 

\begin{table}[ht]
\caption{Comparison of ACC and AUC between benchmark knowledge tracing models and SAINT+.}
\centering
\begin{tabular}{ccc}
\toprule
Model & ACC & AUC \\ 
\midrule
DKT & 0.7060 & 0.7638 \\
DKVMN & 0.7079 & 0.7668 \\
SAKT & 0.7073 & 0.7663 \\
\midrule
SAINT & 0.7178 & 0.7816 \\
SAINT+ & \textbf{0.7252} & \textbf{0.7914} \\
\bottomrule
\end{tabular}
\label{tab:main}
\end{table}

\subsection{Ablation Test}
\subsubsection{Temporal Feature Embedding: Continuous vs. Categorical}
We compare different approaches for embedding temporal features: continuous and categorical.
Modeling elapsed time in continuous (resp. categorical) fashion assumes that the relationship between a student’s knowledge status and the difficulty of a question for the student is a smooth (resp. stepwise) function of time.
Similarly, for lag time, continuous (resp. categorical) modeling addresses various aspects in a students’ learning process including forgetting, re-organizing concepts and improvement change smoothly (resp. discretely) over time.
Table \ref{tab:embedding} show that the best result is obtained when using continuous embedding for elapsed time and categorical embedding for lag time.

\begin{table}[ht]
\caption{Comparison of different approaches for temporal feature embeddings: continuous and categorical for elapsed time and lag time.}
\centering
\begin{tabular}{cccc}
\toprule
Elapsed time & Lag time & ACC & AUC \\ 
\midrule
Continuous & Continuous & 0.7239 & 0.7898 \\
Continuous & Categorical & \textbf{0.7252} & \textbf{0.7914} \\
Categorical & Continuous & 0.7245 & 0.7911 \\
Categorical & Categorical & 0.7235 & 0.7900 \\
\bottomrule
\end{tabular}
\label{tab:embedding}
\end{table}

\subsubsection{Elapsed Time vs. Lag Time}
We verify the contribution of each temporal feature by comparing performance improvements of two variants of SAINT: 1) SAINT with elapsed time only, and 2) SAINT with lag time only.
The results are described in Table \ref{tab:feature}.
When used alone, lag time increases performance more than elapsed time.
Moreover, regardless of the temporal features used, utilizing only one feature is not good as using both features together (SAINT+), while certain improvements are obtained compared to SAINT.

\begin{table}[ht]
\caption{Contribution of each temporal feature.
ET (resp. LT) stands for elapsed time (resp. lag time) and SAINT+ET (resp. SAINT+LT) only uses elapsed time (resp. lag time).}
\centering
\begin{tabular}{ccc}
\toprule
 & ACC & AUC \\ 
\midrule
SAINT & 0.7178 & 0.7816 \\
SAINT + ET & 0.7213 & 0.7858 \\
SAINT + LT & 0.7239 & 0.7898 \\
SAINT + ET + LT & \textbf{0.7252} & \textbf{0.7914} \\
\bottomrule
\end{tabular}
\label{tab:feature}
\end{table}

\subsubsection{Integrating Temporal Features: Encoder vs. Decoder}
The temporal information from the interaction logs, elapsed time and lag time, are provided as decoder features in SAINT+ as described in Figure \ref{fig:model}.
Since both elapsed time and lag time arise as a result of student response, this approach is naturally aligned with the core idea of SAINT that providing exercise information to the encoder and response information to the decoder is appropriate for knowledge tracing.
We compare SAINT+ (decoder only) with other two variants: feeding the temporal features to 1) the encoder only, and 2) both the encoder and the decoder.
Table \ref{tab:encdec} summarizes the results.
As expected, SAINT+ shows the best performance among the variants.
Also, SAINT+ and the variants show better results than SAINT, demonstrating that the temporal features provide useful information for estimating knowledge status.

\begin{table}[ht]
\caption{Comparison of methods for integrating temporal features: encoder only (Enc), decoder only (Dec) and both encoder and decoder (Enc+Dec).}
\centering
\normalsize
\begin{tabular}{ccc}
\toprule
 & ACC & AUC \\ 
\midrule
SAINT & 0.7178 & 0.7816 \\
Enc & 0.7216 & 0.7866 \\
Dec & \textbf{0.7252} & \textbf{0.7914} \\
Enc + Dec & 0.7221 & 0.7872 \\
\bottomrule
\end{tabular}
\label{tab:encdec}
\end{table}

%% file: 6_conclusion.tex
\section{Conclusion}
In this paper, we proposed SAINT+, a Transformer based knowledge tracing model that processes exercise information and student response information separately, and integrates two temporal feature embeddings into the response embeddings: elapsed time and lag time.
Experiments conducted on EdNet dataset show that SAINT+ improves SAINT, the former state-of-the-art knowledge tracing model, in both ACC and AUC.
Furthermore, the best result was obtained by incorporating the temporal features into the decoder input, verifying the hypothesis that separately processing exercise information and student response information is appropriate for knowledge tracing.
Avenues of future work include 1) modeling not only students’ problem-solving records, but also various learning activities, such as watching lectures and studying explanations for each exercise, 2) exploring architectures for knowledge tracing models other than Transformer based encoder-decoder model that separately processes exercise information and student response information.

%% file: main.bbl

\begin{thebibliography}{27}


\ifx \showCODEN    \undefined \def \showCODEN     #1{\unskip}     \fi
\ifx \showDOI      \undefined \def \showDOI       #1{#1}\fi
\ifx \showISBNx    \undefined \def \showISBNx     #1{\unskip}     \fi
\ifx \showISBNxiii \undefined \def \showISBNxiii  #1{\unskip}     \fi
\ifx \showISSN     \undefined \def \showISSN      #1{\unskip}     \fi
\ifx \showLCCN     \undefined \def \showLCCN      #1{\unskip}     \fi
\ifx \shownote     \undefined \def \shownote      #1{#1}          \fi
\ifx \showarticletitle \undefined \def \showarticletitle #1{#1}   \fi
\ifx \showURL      \undefined \def \showURL       {\relax}        \fi
\providecommand\bibfield[2]{#2}
\providecommand\bibinfo[2]{#2}
\providecommand\natexlab[1]{#1}
\providecommand\showeprint[2][]{arXiv:#2}

\bibitem[\protect\citeauthoryear{Ba, Kiros, and Hinton}{Ba
  et~al\mbox{.}}{2016}]%
        {ba2016layer}
\bibfield{author}{\bibinfo{person}{Jimmy~Lei Ba}, \bibinfo{person}{Jamie~Ryan
  Kiros}, {and} \bibinfo{person}{Geoffrey~E Hinton}.}
  \bibinfo{year}{2016}\natexlab{}.
\newblock \showarticletitle{Layer normalization}.
\newblock \bibinfo{journal}{\emph{arXiv preprint arXiv:1607.06450}}
  (\bibinfo{year}{2016}).
\newblock


\bibitem[\protect\citeauthoryear{Choi, Lee, Cho, Baek, Kim, Cha, Shin, Bae, and
  Heo}{Choi et~al\mbox{.}}{2020a}]%
        {choi2020towards}
\bibfield{author}{\bibinfo{person}{Youngduck Choi}, \bibinfo{person}{Youngnam
  Lee}, \bibinfo{person}{Junghyun Cho}, \bibinfo{person}{Jineon Baek},
  \bibinfo{person}{Byungsoo Kim}, \bibinfo{person}{Yeongmin Cha},
  \bibinfo{person}{Dongmin Shin}, \bibinfo{person}{Chan Bae}, {and}
  \bibinfo{person}{Jaewe Heo}.} \bibinfo{year}{2020}\natexlab{a}.
\newblock \showarticletitle{Towards an Appropriate Query, Key, and Value
  Computation for Knowledge Tracing}.
\newblock \bibinfo{journal}{\emph{arXiv preprint arXiv:2002.07033}}
  (\bibinfo{year}{2020}).
\newblock


\bibitem[\protect\citeauthoryear{Choi, Lee, Shin, Cho, Park, Lee, Baek, Bae,
  Kim, and Heo}{Choi et~al\mbox{.}}{2020b}]%
        {choi2020ednet}
\bibfield{author}{\bibinfo{person}{Youngduck Choi}, \bibinfo{person}{Youngnam
  Lee}, \bibinfo{person}{Dongmin Shin}, \bibinfo{person}{Junghyun Cho},
  \bibinfo{person}{Seoyon Park}, \bibinfo{person}{Seewoo Lee},
  \bibinfo{person}{Jineon Baek}, \bibinfo{person}{Chan Bae},
  \bibinfo{person}{Byungsoo Kim}, {and} \bibinfo{person}{Jaewe Heo}.}
  \bibinfo{year}{2020}\natexlab{b}.
\newblock \showarticletitle{Ednet: A large-scale hierarchical dataset in
  education}. In \bibinfo{booktitle}{\emph{International Conference on
  Artificial Intelligence in Education}}. Springer, \bibinfo{pages}{69--73}.
\newblock


\bibitem[\protect\citeauthoryear{Corbett and Anderson}{Corbett and
  Anderson}{1994}]%
        {corbertt_1994}
\bibfield{author}{\bibinfo{person}{Albert~T Corbett} {and}
  \bibinfo{person}{John~R Anderson}.} \bibinfo{year}{1994}\natexlab{}.
\newblock \showarticletitle{Knowledge tracing: Modeling the acquisition of
  procedural knowledge}.
\newblock \bibinfo{journal}{\emph{User modeling and user-adapted interaction}}
  \bibinfo{volume}{4}, \bibinfo{number}{4} (\bibinfo{year}{1994}),
  \bibinfo{pages}{253--278}.
\newblock


\bibitem[\protect\citeauthoryear{d~Baker, Corbett, and Aleven}{d~Baker
  et~al\mbox{.}}{2008}]%
        {d2008more}
\bibfield{author}{\bibinfo{person}{Ryan~SJ d Baker}, \bibinfo{person}{Albert~T
  Corbett}, {and} \bibinfo{person}{Vincent Aleven}.}
  \bibinfo{year}{2008}\natexlab{}.
\newblock \showarticletitle{More accurate student modeling through contextual
  estimation of slip and guess probabilities in bayesian knowledge tracing}. In
  \bibinfo{booktitle}{\emph{International conference on intelligent tutoring
  systems}}. Springer, \bibinfo{pages}{406--415}.
\newblock


\bibitem[\protect\citeauthoryear{Ghosh, Heffernan, and Lan}{Ghosh
  et~al\mbox{.}}{2020}]%
        {ghosh2020context}
\bibfield{author}{\bibinfo{person}{Aritra Ghosh}, \bibinfo{person}{Neil
  Heffernan}, {and} \bibinfo{person}{Andrew~S Lan}.}
  \bibinfo{year}{2020}\natexlab{}.
\newblock \showarticletitle{Context-aware attentive knowledge tracing}. In
  \bibinfo{booktitle}{\emph{Proceedings of the 26th ACM SIGKDD International
  Conference on Knowledge Discovery \& Data Mining}}.
  \bibinfo{pages}{2330--2339}.
\newblock


\bibitem[\protect\citeauthoryear{He, Zhang, Ren, and Sun}{He
  et~al\mbox{.}}{2016}]%
        {he2016deep}
\bibfield{author}{\bibinfo{person}{Kaiming He}, \bibinfo{person}{Xiangyu
  Zhang}, \bibinfo{person}{Shaoqing Ren}, {and} \bibinfo{person}{Jian Sun}.}
  \bibinfo{year}{2016}\natexlab{}.
\newblock \showarticletitle{Deep residual learning for image recognition}. In
  \bibinfo{booktitle}{\emph{Proceedings of the IEEE conference on computer
  vision and pattern recognition}}. \bibinfo{pages}{770--778}.
\newblock


\bibitem[\protect\citeauthoryear{Hochreiter and Schmidhuber}{Hochreiter and
  Schmidhuber}{1997}]%
        {hochreiter1997long}
\bibfield{author}{\bibinfo{person}{Sepp Hochreiter} {and}
  \bibinfo{person}{J{\"u}rgen Schmidhuber}.} \bibinfo{year}{1997}\natexlab{}.
\newblock \showarticletitle{Long short-term memory}.
\newblock \bibinfo{journal}{\emph{Neural computation}} \bibinfo{volume}{9},
  \bibinfo{number}{8} (\bibinfo{year}{1997}), \bibinfo{pages}{1735--1780}.
\newblock


\bibitem[\protect\citeauthoryear{Huang, Yin, Chen, Xiong, Su, Hu,
  et~al\mbox{.}}{Huang et~al\mbox{.}}{2019}]%
        {liu_2019}
\bibfield{author}{\bibinfo{person}{Zhenya Huang}, \bibinfo{person}{Yu Yin},
  \bibinfo{person}{Enhong Chen}, \bibinfo{person}{Hui Xiong},
  \bibinfo{person}{Yu Su}, \bibinfo{person}{Guoping Hu}, {et~al\mbox{.}}}
  \bibinfo{year}{2019}\natexlab{}.
\newblock \showarticletitle{EKT: Exercise-aware Knowledge Tracing for Student
  Performance Prediction}.
\newblock \bibinfo{journal}{\emph{IEEE Transactions on Knowledge and Data
  Engineering}} (\bibinfo{year}{2019}).
\newblock


\bibitem[\protect\citeauthoryear{Kasurinen and Nikula}{Kasurinen and
  Nikula}{2009}]%
        {kasurinen2009estimating}
\bibfield{author}{\bibinfo{person}{Jussi Kasurinen} {and}
  \bibinfo{person}{Uolevi Nikula}.} \bibinfo{year}{2009}\natexlab{}.
\newblock \showarticletitle{Estimating programming knowledge with Bayesian
  knowledge tracing}.
\newblock \bibinfo{journal}{\emph{ACM SIGCSE Bulletin}} \bibinfo{volume}{41},
  \bibinfo{number}{3} (\bibinfo{year}{2009}), \bibinfo{pages}{313--317}.
\newblock


\bibitem[\protect\citeauthoryear{Lee, Chung, Cha, and Suh}{Lee
  et~al\mbox{.}}{2016}]%
        {lee2016machine}
\bibfield{author}{\bibinfo{person}{Kangwook Lee}, \bibinfo{person}{Jichan
  Chung}, \bibinfo{person}{Yeongmin Cha}, {and} \bibinfo{person}{Changho Suh}.}
  \bibinfo{year}{2016}\natexlab{}.
\newblock \showarticletitle{Machine Learning Approaches for Learning Analytics:
  Collaborative Filtering Or Regression With Experts?}. In
  \bibinfo{booktitle}{\emph{NIPS Workshop, Dec}}. \bibinfo{pages}{1--11}.
\newblock


\bibitem[\protect\citeauthoryear{Lee, Choi, Cho, Fabbri, Loh, Hwang, Lee, Kim,
  and Radev}{Lee et~al\mbox{.}}{2019}]%
        {lee_2019}
\bibfield{author}{\bibinfo{person}{Youngnam Lee}, \bibinfo{person}{Youngduck
  Choi}, \bibinfo{person}{Junghyun Cho}, \bibinfo{person}{Alexander~R Fabbri},
  \bibinfo{person}{Hyunbin Loh}, \bibinfo{person}{Chanyou Hwang},
  \bibinfo{person}{Yongku Lee}, \bibinfo{person}{Sang-Wook Kim}, {and}
  \bibinfo{person}{Dragomir Radev}.} \bibinfo{year}{2019}\natexlab{}.
\newblock \showarticletitle{Creating A Neural Pedagogical Agent by Jointly
  Learning to Review and Assess}.
\newblock \bibinfo{journal}{\emph{arXiv preprint arXiv:1906.10910}}
  (\bibinfo{year}{2019}).
\newblock


\bibitem[\protect\citeauthoryear{Nakagawa, Iwasawa, and Matsuo}{Nakagawa
  et~al\mbox{.}}{2019}]%
        {nakagawa2019graph}
\bibfield{author}{\bibinfo{person}{Hiromi Nakagawa}, \bibinfo{person}{Yusuke
  Iwasawa}, {and} \bibinfo{person}{Yutaka Matsuo}.}
  \bibinfo{year}{2019}\natexlab{}.
\newblock \showarticletitle{Graph-based Knowledge Tracing: Modeling Student
  Proficiency Using Graph Neural Network}. In \bibinfo{booktitle}{\emph{2019
  IEEE/WIC/ACM International Conference on Web Intelligence (WI)}}. IEEE,
  \bibinfo{pages}{156--163}.
\newblock


\bibitem[\protect\citeauthoryear{Pandey and Karypis}{Pandey and
  Karypis}{2019}]%
        {pandy_2019}
\bibfield{author}{\bibinfo{person}{Shalini Pandey} {and}
  \bibinfo{person}{George Karypis}.} \bibinfo{year}{2019}\natexlab{}.
\newblock \showarticletitle{A Self-Attentive model for Knowledge Tracing}.
\newblock \bibinfo{journal}{\emph{arXiv preprint arXiv:1907.06837}}
  (\bibinfo{year}{2019}).
\newblock


\bibitem[\protect\citeauthoryear{Pardos and Heffernan}{Pardos and
  Heffernan}{2010a}]%
        {pardos2010navigating}
\bibfield{author}{\bibinfo{person}{Zachary Pardos} {and} \bibinfo{person}{Neil
  Heffernan}.} \bibinfo{year}{2010}\natexlab{a}.
\newblock \showarticletitle{Navigating the parameter space of Bayesian
  Knowledge Tracing models: Visualizations of the convergence of the
  Expectation Maximization algorithm}. In \bibinfo{booktitle}{\emph{Educational
  Data Mining 2010}}.
\newblock


\bibitem[\protect\citeauthoryear{Pardos, Bergner, Seaton, and Pritchard}{Pardos
  et~al\mbox{.}}{2013}]%
        {pardos2013adapting}
\bibfield{author}{\bibinfo{person}{Zachary~A Pardos}, \bibinfo{person}{Yoav
  Bergner}, \bibinfo{person}{Daniel~T Seaton}, {and} \bibinfo{person}{David~E
  Pritchard}.} \bibinfo{year}{2013}\natexlab{}.
\newblock \showarticletitle{Adapting Bayesian Knowledge Tracing to a Massive
  Open Online Course in edX.}
\newblock \bibinfo{journal}{\emph{EDM}}  \bibinfo{volume}{13}
  (\bibinfo{year}{2013}), \bibinfo{pages}{137--144}.
\newblock


\bibitem[\protect\citeauthoryear{Pardos and Heffernan}{Pardos and
  Heffernan}{2010b}]%
        {pardos2010modeling}
\bibfield{author}{\bibinfo{person}{Zachary~A Pardos} {and}
  \bibinfo{person}{Neil~T Heffernan}.} \bibinfo{year}{2010}\natexlab{b}.
\newblock \showarticletitle{Modeling individualization in a bayesian networks
  implementation of knowledge tracing}. In
  \bibinfo{booktitle}{\emph{International Conference on User Modeling,
  Adaptation, and Personalization}}. Springer, \bibinfo{pages}{255--266}.
\newblock


\bibitem[\protect\citeauthoryear{Piech, Bassen, Huang, Ganguli, Sahami, Guibas,
  and Sohl-Dickstein}{Piech et~al\mbox{.}}{2015}]%
        {piech_2015}
\bibfield{author}{\bibinfo{person}{Chris Piech}, \bibinfo{person}{Jonathan
  Bassen}, \bibinfo{person}{Jonathan Huang}, \bibinfo{person}{Surya Ganguli},
  \bibinfo{person}{Mehran Sahami}, \bibinfo{person}{Leonidas~J Guibas}, {and}
  \bibinfo{person}{Jascha Sohl-Dickstein}.} \bibinfo{year}{2015}\natexlab{}.
\newblock \showarticletitle{Deep knowledge tracing}. In
  \bibinfo{booktitle}{\emph{Advances in neural information processing
  systems}}. \bibinfo{pages}{505--513}.
\newblock


\bibitem[\protect\citeauthoryear{Qiu, Qi, Lu, Pardos, and Heffernan}{Qiu
  et~al\mbox{.}}{2011}]%
        {qiu2011does}
\bibfield{author}{\bibinfo{person}{Yumeng Qiu}, \bibinfo{person}{Yingmei Qi},
  \bibinfo{person}{Hanyuan Lu}, \bibinfo{person}{Zachary~A Pardos}, {and}
  \bibinfo{person}{Neil~T Heffernan}.} \bibinfo{year}{2011}\natexlab{}.
\newblock \showarticletitle{Does Time Matter? Modeling the Effect of Time with
  Bayesian Knowledge Tracing.}. In \bibinfo{booktitle}{\emph{EDM}}.
  \bibinfo{pages}{139--148}.
\newblock


\bibitem[\protect\citeauthoryear{Sao~Pedro, Baker, and Gobert}{Sao~Pedro
  et~al\mbox{.}}{2013}]%
        {sao2013incorporating}
\bibfield{author}{\bibinfo{person}{Michael Sao~Pedro}, \bibinfo{person}{Ryan
  Baker}, {and} \bibinfo{person}{Janice Gobert}.}
  \bibinfo{year}{2013}\natexlab{}.
\newblock \showarticletitle{Incorporating scaffolding and tutor context into
  bayesian knowledge tracing to predict inquiry skill acquisition}. In
  \bibinfo{booktitle}{\emph{Educational Data Mining 2013}}.
\newblock


\bibitem[\protect\citeauthoryear{Shen, Liu, Chen, Wu, Huang, Zhao, Su, Ma, and
  Wang}{Shen et~al\mbox{.}}{2020}]%
        {shen2020convolutional}
\bibfield{author}{\bibinfo{person}{Shuanghong Shen}, \bibinfo{person}{Qi Liu},
  \bibinfo{person}{Enhong Chen}, \bibinfo{person}{Han Wu},
  \bibinfo{person}{Zhenya Huang}, \bibinfo{person}{Weihao Zhao},
  \bibinfo{person}{Yu Su}, \bibinfo{person}{Haiping Ma}, {and}
  \bibinfo{person}{Shijin Wang}.} \bibinfo{year}{2020}\natexlab{}.
\newblock \showarticletitle{Convolutional Knowledge Tracing: Modeling
  Individualization in Student Learning Process}. In
  \bibinfo{booktitle}{\emph{Proceedings of the 43rd International ACM SIGIR
  Conference on Research and Development in Information Retrieval}}.
  \bibinfo{pages}{1857--1860}.
\newblock


\bibitem[\protect\citeauthoryear{Thai-Nghe, Drumond, Krohn-Grimberghe, and
  Schmidt-Thieme}{Thai-Nghe et~al\mbox{.}}{2010}]%
        {thai2010recommender}
\bibfield{author}{\bibinfo{person}{Nguyen Thai-Nghe}, \bibinfo{person}{Lucas
  Drumond}, \bibinfo{person}{Artus Krohn-Grimberghe}, {and}
  \bibinfo{person}{Lars Schmidt-Thieme}.} \bibinfo{year}{2010}\natexlab{}.
\newblock \showarticletitle{Recommender system for predicting student
  performance}.
\newblock \bibinfo{journal}{\emph{Procedia Computer Science}}
  \bibinfo{volume}{1}, \bibinfo{number}{2} (\bibinfo{year}{2010}),
  \bibinfo{pages}{2811--2819}.
\newblock


\bibitem[\protect\citeauthoryear{Tong, Zhou, and Wang}{Tong
  et~al\mbox{.}}{2020}]%
        {tong2020hgkt}
\bibfield{author}{\bibinfo{person}{Hanshuang Tong}, \bibinfo{person}{Yun Zhou},
  {and} \bibinfo{person}{Zhen Wang}.} \bibinfo{year}{2020}\natexlab{}.
\newblock \showarticletitle{HGKT: Introducing Problem Schema with Hierarchical
  Exercise Graph for Knowledge Tracing}.
\newblock \bibinfo{journal}{\emph{arXiv preprint arXiv:2006.16915}}
  (\bibinfo{year}{2020}).
\newblock


\bibitem[\protect\citeauthoryear{van De~Sande}{van De~Sande}{2013}]%
        {van2013properties}
\bibfield{author}{\bibinfo{person}{Brett van De~Sande}.}
  \bibinfo{year}{2013}\natexlab{}.
\newblock \showarticletitle{Properties of the bayesian knowledge tracing
  model}.
\newblock \bibinfo{journal}{\emph{Journal of Educational Data Mining}}
  \bibinfo{volume}{5}, \bibinfo{number}{2} (\bibinfo{year}{2013}),
  \bibinfo{pages}{1}.
\newblock


\bibitem[\protect\citeauthoryear{Vaswani, Shazeer, Parmar, Uszkoreit, Jones,
  Gomez, Kaiser, and Polosukhin}{Vaswani et~al\mbox{.}}{2017}]%
        {vaswani_2017}
\bibfield{author}{\bibinfo{person}{Ashish Vaswani}, \bibinfo{person}{Noam
  Shazeer}, \bibinfo{person}{Niki Parmar}, \bibinfo{person}{Jakob Uszkoreit},
  \bibinfo{person}{Llion Jones}, \bibinfo{person}{Aidan~N Gomez},
  \bibinfo{person}{{\L}ukasz Kaiser}, {and} \bibinfo{person}{Illia
  Polosukhin}.} \bibinfo{year}{2017}\natexlab{}.
\newblock \showarticletitle{Attention is all you need}. In
  \bibinfo{booktitle}{\emph{Advances in neural information processing
  systems}}. \bibinfo{pages}{5998--6008}.
\newblock


\bibitem[\protect\citeauthoryear{Yudelson, Koedinger, and Gordon}{Yudelson
  et~al\mbox{.}}{2013}]%
        {yudelson2013individualized}
\bibfield{author}{\bibinfo{person}{Michael~V Yudelson},
  \bibinfo{person}{Kenneth~R Koedinger}, {and} \bibinfo{person}{Geoffrey~J
  Gordon}.} \bibinfo{year}{2013}\natexlab{}.
\newblock \showarticletitle{Individualized bayesian knowledge tracing models}.
  In \bibinfo{booktitle}{\emph{International conference on artificial
  intelligence in education}}. Springer, \bibinfo{pages}{171--180}.
\newblock


\bibitem[\protect\citeauthoryear{Zhang, Shi, King, and Yeung}{Zhang
  et~al\mbox{.}}{2017}]%
        {zhang_2017}
\bibfield{author}{\bibinfo{person}{Jiani Zhang}, \bibinfo{person}{Xingjian
  Shi}, \bibinfo{person}{Irwin King}, {and} \bibinfo{person}{Dit-Yan Yeung}.}
  \bibinfo{year}{2017}\natexlab{}.
\newblock \showarticletitle{Dynamic key-value memory networks for knowledge
  tracing}. In \bibinfo{booktitle}{\emph{Proceedings of the 26th international
  conference on World Wide Web}}. International World Wide Web Conferences
  Steering Committee, \bibinfo{pages}{765--774}.
\newblock


\end{thebibliography}
